%% file: iclr2027_conference.tex
\documentclass{article} 
\usepackage{iclr2027_conference,times}

\input{math_commands.tex}

\usepackage{hyperref}
\usepackage{url}
\usepackage{booktabs}        
\usepackage{threeparttable} 
\usepackage{tabularx}
\usepackage{array}
\usepackage{graphicx}
\usepackage{subcaption}
\usepackage{caption}  
\usepackage[utf8]{inputenc}
\usepackage{CJKutf8}
\usepackage{needspace}
\usepackage{booktabs}
\usepackage{array}
\usepackage{tabularx}
\usepackage[table]{xcolor} 
\usepackage{booktabs}
\usepackage{fontawesome5}
\usepackage{wrapfig}
\usepackage{xcolor}
\usepackage{fontawesome5}
\usepackage{hyperref}
\usepackage{tikz} 
\definecolor{myblue}{RGB}{100, 149, 237} 

\title{KuaiSearch: An E-Commerce Search Dataset with Authentic Queries
and Product Texts for Recall, Ranking, and Relevance}

\author{
Yupeng Li$^{1,\ast,\ddagger}$ \quad
Ben Chen$^{2,\ast}$ \quad
Mingyue Cheng$^{1,\dagger}$ \quad
Qingpeng Cai$^{2}$ \quad
Zhiding Liu$^{1}$
\\[0.35em]
\bfseries
Xuxin Zhang$^{2}$ \quad
Zhentao Song$^{2}$ \quad
Chenyi Lei$^{2,\dagger}$ \quad
Qi Liu$^{1}$ \quad
Peng Jiang$^{2}$
\\[0.7em]
\normalfont\small
$^{1}$State Key Laboratory of Cognitive Intelligence, University of Science and Technology of China
\\[0.15em]
$^{2}$Kuaishou Technology
}

\makeatletter
\g@addto@macro\@thanks{%
  \footnotetext[1]{Equal contribution.}%
  \footnotetext[2]{Corresponding authors.}%
  \footnotetext[3]{Work done during internship at Kuaishou Technology.}%
}
\makeatother

\iclrfinalcopy 
\begin{document}

\maketitle

\begin{abstract}
E-commerce search connects user needs with massive product inventories, yet real-world systems face challenges from ambiguous queries, noisy product texts, and diverse user preferences. Recent advances in large language models (LLMs) offer new opportunities for semantic understanding and intent modeling, but existing e-commerce search datasets remain limited by heuristically constructed queries, popularity-based filtering, anonymized texts, and incomplete coverage of the search pipeline. These limitations hinder realistic evaluation of LLM-based e-commerce search. To address this gap, we introduce \textbf{KuaiSearch}, a large-scale dataset built from real user search interactions on the Kuaishou platform. KuaiSearch preserves authentic queries and natural-language product texts, retains cold-start users and long-tail products, and covers three key stages of the search pipeline: \emph{recall}, \emph{ranking}, and \emph{relevance judgment}. We further provide comprehensive analyses of users, products, and queries, together with benchmark experiments on representative search tasks. Results demonstrate the value of KuaiSearch as a realistic benchmark for e-commerce search research. The code is available at \url{https://github.com/benchen4395/KuaiSearch}.
\end{abstract}

\section{Introduction}
E-commerce search plays a vital role in our daily lives by serving as the primary interface through which users express information needs, explore product content, and make purchasing decisions on online platforms \citep{survey1,survey2}. As the scale and diversity of online marketplaces continue to grow, e-commerce search has become a critical infrastructure component that directly influences user satisfaction, engagement, and conversion behavior, and largely determines the overall effectiveness of modern e-commerce systems. As a result, search quality is crucial for effectively connecting user demand with massive product inventories.

In e-commerce scenarios, user queries are often broad and ambiguous, typically consisting of brand names, product categories, or attribute-related keywords rather than well-specified product descriptions \citep{queryvague1,queryvague2}. Meanwhile, product-side texts are generally lengthy and noisy, as merchants may insert irrelevant or weakly related keywords to increase product exposure. Consequently, core information such as brands, attributes, and categories is often presented in an unstructured manner, weakening the discriminative capability of representation models \citep{titlevague1,titlevague2}. In addition, e-commerce search exhibits substantial user heterogeneity, such that the same query may correspond to different budgets, preferences, and usage contexts. Effective search systems must therefore capture both query semantics and personalized user intent \citep{per1,per2,per3,per4}. Under these conditions, static relevance modeling between queries and product texts remains insufficient. Recent advances in large language models provide a promising direction for bridging this semantic gap through enhanced user profiling, preference modeling, and semantic understanding \citep{onesearch,onevision,zhiding}.

\begin{table}[t]
\centering
\caption{Statistical comparison of e-commerce search datasets.}  
\label{tab:dataset_statistics}
\begin{threeparttable}
\begin{tabular}{lcccc}
\toprule
\textbf{Dataset}
& \textbf{\# Users}
& \textbf{\# Items}
& \textbf{\# Queries}
& \textbf{Text form} \\
\midrule
Amazon
& 192,403
& 63,001
& 3,221$^{*}$
& text \\

Amazon-C4
& 20,838
& 1,058,417
& 21,223$^{*}$
& text \\

JDsearch
& 173,831
& 12,872,736
& 171,728
& anon'd \\
KuaiSearch-Lite
& 102,086
& 6,634,118
& 555,553
& text \\
KuaiSearch
& 331,930
& 18,605,582
& 2,574,949
& text \\
\bottomrule
\end{tabular}

\begin{tablenotes}
\footnotesize
\item[*] Queries are constructed rather than collected from real user searches
\end{tablenotes}
\end{threeparttable}
\vspace{-0.1in}
\end{table}

However, existing e-commerce search datasets remain limited in data scale, textual completeness, behavioral signals, and search-task coverage. Early studies have relied on Amazon datasets containing heuristically constructed queries, while filtering cold-start users and long-tail products \citep{amazonquery,per3,amazonc4}. Such preprocessing produces data distributions that differ from real-world search scenarios. JDsearch provides real queries and multiple user--product interactions while preserving cold-start users and long-tail products \citep{jdsearch}; nevertheless, its anonymized textual content limits research on LLM-based search. More recently, Qilin provides natural-language text, but its setting is closer to content browsing, covers only one day of interactions, and lacks sufficient behavioral signals for modeling purchase intent \citep{qilin}. Furthermore, most existing datasets focus on a single task, such as recall or ranking, making it difficult to evaluate multiple components within a shared real-world domain. Consequently, they remain insufficient for systematic research on real-world LLM-based e-commerce search.

To address the aforementioned challenges, we construct and release \textbf{KuaiSearch}, an e-commerce search dataset built upon one of the most popular short-video platforms in China, Kuaishou. KuaiSearch explicitly covers three key tasks in the e-commerce search pipeline---\emph{recall}, \emph{ranking}, and \emph{relevance judgment}---and supports the evaluation of representative methods for each task within a shared e-commerce domain. The dataset consists of approximately 330,000 users, 18,000,000 products, and 2,500,000 real search queries, providing a large-scale benchmark for real-world e-commerce search research. During dataset construction, we fully retain real user search queries without applying any popularity-based filtering on products, and include users with diverse historical behavior lengths, thus preserving data distributions that closely reflect commercial environments. We further conduct multi-perspective analyses and benchmark representative models on KuaiSearch, demonstrating its research value and practical utility for e-commerce search.

In summary, the main contributions of this work are as follows:
\begin{itemize}
\item We release KuaiSearch, a large-scale dataset of real queries,
full-category products, and long-term user behavior that preserves
realistic e-commerce search distributions.

\item KuaiSearch supports three key e-commerce search tasks:
\emph{recall}, \emph{ranking}, and \emph{relevance judgment}, providing
a benchmark for representative methods across different tasks.

\item We provide comprehensive dataset analysis and benchmark representative models on KuaiSearch, demonstrating its effectiveness and value for e-commerce search research.
\end{itemize}

\section{Related Work}
\subsection{E-commerce Search}
Early e-commerce search systems mainly rely on traditional information retrieval methods. Sparse retrieval approaches such as TF-IDF and BM25 \citep{tf-idf,bm25} remain strong baselines due to their efficiency, while dense retrieval models such as DPR and ANCE \citep{dpr,ance} improve semantic matching by encoding queries and products into dense representations. Neural ranking models \citep{per1,per2,per3} further re-rank retrieved candidates with richer features for more fine-grained and personalized ranking. More recently, generative retrieval has emerged as a promising paradigm. Early works including DSI, DSI-QG, and LTRGR \citep{dsi,dsi-qg,ltrgr} establish the foundation of generative retrieval, while OneSearch and OneSug \citep{onesearch,onesug} explore unified generative modeling for e-commerce retrieval and ranking, and OneVision \citep{onevision} extends it to multimodal search. However, existing research is still constrained by the lack of large-scale e-commerce search datasets that jointly contain natural-language queries and authentic user behavior, limiting systematic evaluation in realistic scenarios.

\subsection{Existing Datasets}
Datasets are essential for training and evaluating retrieval systems. General-purpose benchmarks such as MS MARCO and NQ \citep{msmarco,nq} mainly contain explicit question-style queries and lack user behavioral context, making them less suitable for personalized e-commerce search, where queries are often short and ambiguous, product texts are noisy, and user behaviors exhibit strong personalization and long-tail characteristics. Existing e-commerce datasets also have notable limitations. Amazon-based benchmarks \citep{amazonquery,per3,amazonc4} often construct queries heuristically from product metadata and filter cold-start users or long-tail products, leading to distribution shifts from real-world search. JDsearch \citep{jdsearch} provides real queries and user–product interactions while preserving long-tail data, but anonymizes textual content, limiting LLM-based research. Qilin \citep{qilin} retains plaintext content, yet focuses more on content retrieval and contains only short-term interaction data, making it insufficient for modeling complex and long-term e-commerce search behaviors.

\section{The KuaiSearch Dataset}
In this section, we introduce the KuaiSearch dataset, including its construction pipeline and data schema. KuaiSearch is built from real user search data collected from the e-commerce module of Kuaishou, covering diverse real-world search behaviors. The construction process consists of five components: user behavior collection, product metadata collection, user metadata collection, ranking data construction, and relevance data construction. Data from different components can be linked via shared session IDs for integrated analysis across components.

\begin{table}[t]
\centering
\caption{Field Descriptions and Data Sizes of Different Components in KuaiSearch.}
\label{tab:kuaisearch_fields}
\setlength{\tabcolsep}{4pt}
\renewcommand{\arraystretch}{1.05}
\begin{tabularx}{\columnwidth}{@{}l c >{\raggedright\arraybackslash}X@{}}
\hline
\textbf{Table} & \textbf{Size} & \textbf{Fields} \\
\hline
User &
331,930  &
user id, gender, age, location \\
\hline
Item &
18,605,582 &
item id, title, brand id/name, seller id/name,
category L1 id/name, category L2 id/name, category L3 id/name \\
\hline
Recall &
2,574,949 &
user id, session id, time index, query, impressed item ids, clicked item ids, purchased item ids \\
\hline
Ranking &
81,401,477  &
user id, user statistical features, fan number, follow number, session id, time index, query, search entrance, recently clicked item ids, recently purchased item ids, target item id, target item statistical features, target item price,
is clicked, is purchased \\
\hline
Relevance &
46,422 &
query, title, brand name, seller name, attribute,\allowbreak
score \\
\hline

\end{tabularx}
\end{table}

\paragraph{User Behavior Collection.}
We randomly sampled 331,930 users who exhibited core interaction behaviors during the two-month period starting on June 1, 2025. For each user, we collected all product search activities within the corresponding time window. For each search request, we retained the complete set of exposed items and recorded feedback signals, including clicks and purchases. Notably, we also retained search requests with no clicks or purchases to ensure the completeness of user behavior logs. During data construction, no constraints were imposed on the user population or the length of users’ historical behaviors; as a result, the dataset naturally covers diverse user behavior patterns and more faithfully reflects real-world e-commerce search scenarios.

\paragraph{Product Metadata Collection.}
On the product side, we collect metadata, including titles, three-level category hierarchies, brand names, and seller names. All fields are retained in plaintext, enabling research on natural language understanding and semantic matching in e-commerce search.

\paragraph{User Metadata Collection.}
On the user side, we record basic demographic attributes, including gender, age group, and geographic location. These features provide auxiliary signals for modeling user heterogeneity and personalized search behavior.

\paragraph{Ranking Data Construction.}
Based on the complete user interaction logs, we further construct ranking data for each search request. Specifically, we augment the dataset with rich user-level and product-level features. On the user side, these include the numbers of followees and followers, as well as clicks, exposures, purchases, and gross merchandise value (GMV) over the past 30 days. On the product side, we provide exposure count, click count, purchase count, and price. In addition, the ranking data explicitly record the search entrance to characterize differences in user behavior patterns and search intent across scenarios. Furthermore, we include the user’s most recent 20 clicked products and most recent 20 purchased products as sequential behavior features, enabling more fine-grained modeling of preference evolution and intent dynamics.

\paragraph{Relevance Data Construction.}
To support research on relevance modeling, we construct a relevance dataset consisting of 46,422 query–product pairs collected from real user search scenarios. On the product side, we provide product titles, seller names, brand names, and attribute values. All query–product pairs are manually annotated by domain experts with e-commerce expertise using a four-level graded relevance scheme, thereby providing fine-grained supervision signals. Detailed annotation procedures and quality control measures are given in Appendix \ref{sec:annotation_quality}.

KuaiSearch contains real-world e-commerce search interactions from 331,930 users. For privacy and compliance, identifiers are remapped and timestamps are represented using relative values (Appendix~\ref{app:privacy}). We also release KuaiSearch-Lite for efficient validation, ablation studies, and resource-constrained research (Appendix~\ref{app:lite_construction}). Table~\ref{tab:dataset_statistics} compares both versions with existing datasets, while Table~\ref{tab:kuaisearch_fields} and Section~\ref{sec:relevance} summarize the data fields and relevance statistics, respectively.

\section{Primary Data Analysis}

This section presents a primary analysis of KuaiSearch, including (1) search interaction and content analysis over products, users, and queries, and (2) relevance data analysis.

\subsection{Search Interaction and Content Analysis}
In this section, we analyze e-commerce search characteristics in KuaiSearch from the perspectives of products, users, and queries, covering behavioral patterns, content properties, and distributions.

\begin{figure}[t]
    \centering
    \begin{minipage}[t]{0.43\linewidth}
        \centering
        \includegraphics[width=\linewidth]{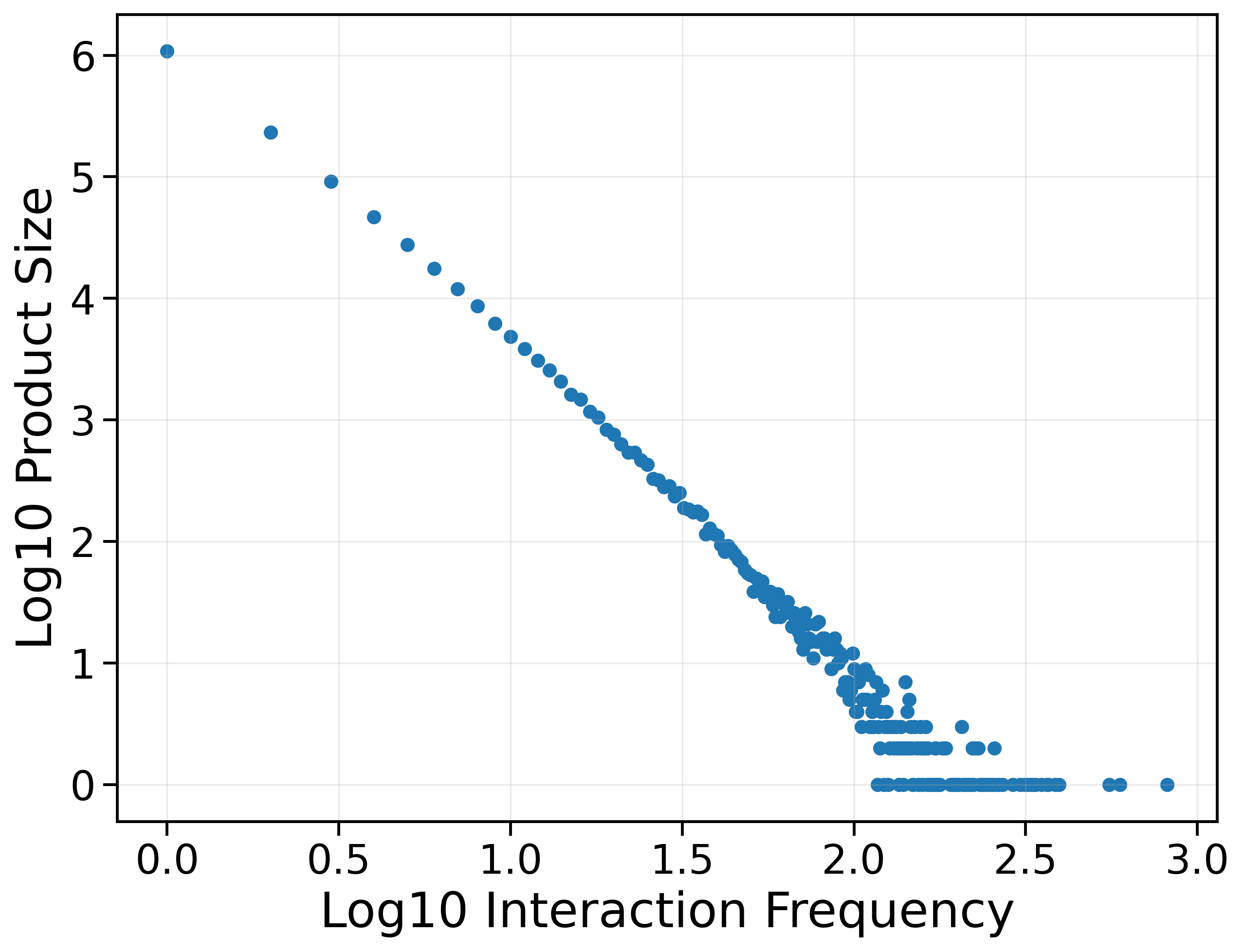}
        \\[-1.5mm]
        {\small (a) Product interaction frequency}
    \end{minipage}
    \hspace{0.04\linewidth}
    \begin{minipage}[t]{0.43\linewidth}
        \centering
        \includegraphics[width=\linewidth]{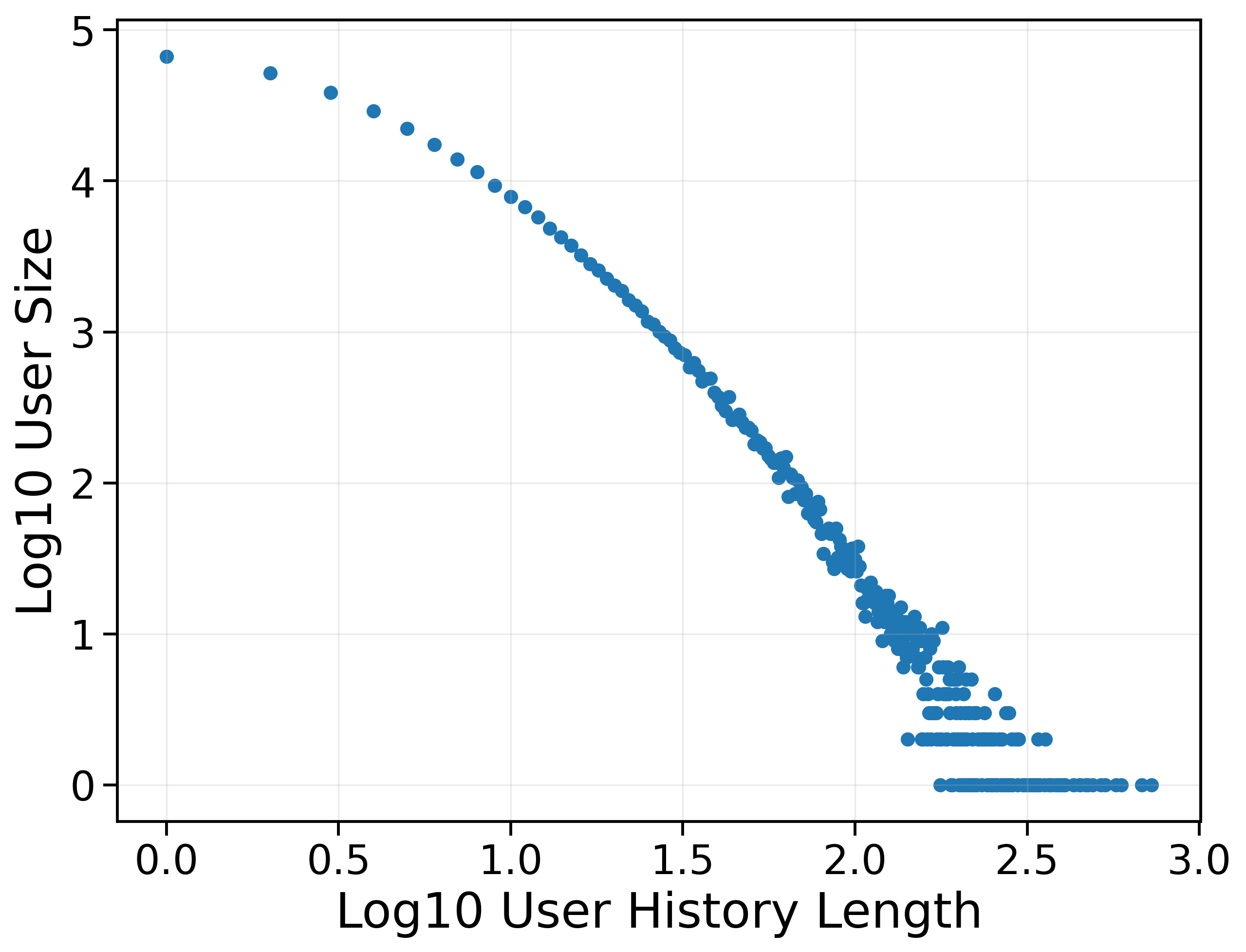}
        \\[-1.5mm]
        {\small (b) User history search frequency}
    \end{minipage}
    \caption{Log-log distributions of product interaction frequencies and searches per user.}
    \label{fig:two_dists}
\end{figure}

\subsubsection{Product Analysis}
First, we present the distribution of product interaction frequencies in the KuaiSearch dataset in Figure~\ref{fig:two_dists} (a). As shown, the interaction frequencies exhibit a pronounced long-tail distribution: a large number of products are interacted with only a few times by users (i.e., cold products), while only a small fraction of products receive frequent interactions. This pronounced long-tail distribution reflects the highly imbalanced popularity distribution of products in real-world e-commerce search scenarios. On the one hand, the vast majority of cold products constitute the main body of the product space, leading to severe data sparsity; on the other hand, a limited number of frequently interacted products concentrate most of the user behavior signals.

\subsubsection{User History Analysis}

Next, we analyze user behavior patterns. Figure~\ref{fig:two_dists} (b) shows the frequency distribution of the number of searches per user in the KuaiSearch dataset. We observe that the dataset includes users with very limited search activity, as well as users with extremely rich search histories. This distribution indicates substantial heterogeneity in user shopping behavior: some users conduct only a few product searches, whereas others exhibit frequent search activity. Consequently, this characteristic creates opportunities for designing differentiated personalization strategies tailored to users with diverse behavioral patterns. For example, when user histories are sparse, models may rely more heavily on general preferences learned from the entire user population; in contrast, for users with extensive search histories, greater emphasis can be placed on modeling individual interests. Such diversity also provides a realistic setting for evaluating the robustness of personalized search models under varying levels of behavioral sparsity.

\begin{wrapfigure}{r}{0.5\textwidth}
    \centering
    \includegraphics[width=\linewidth]
        {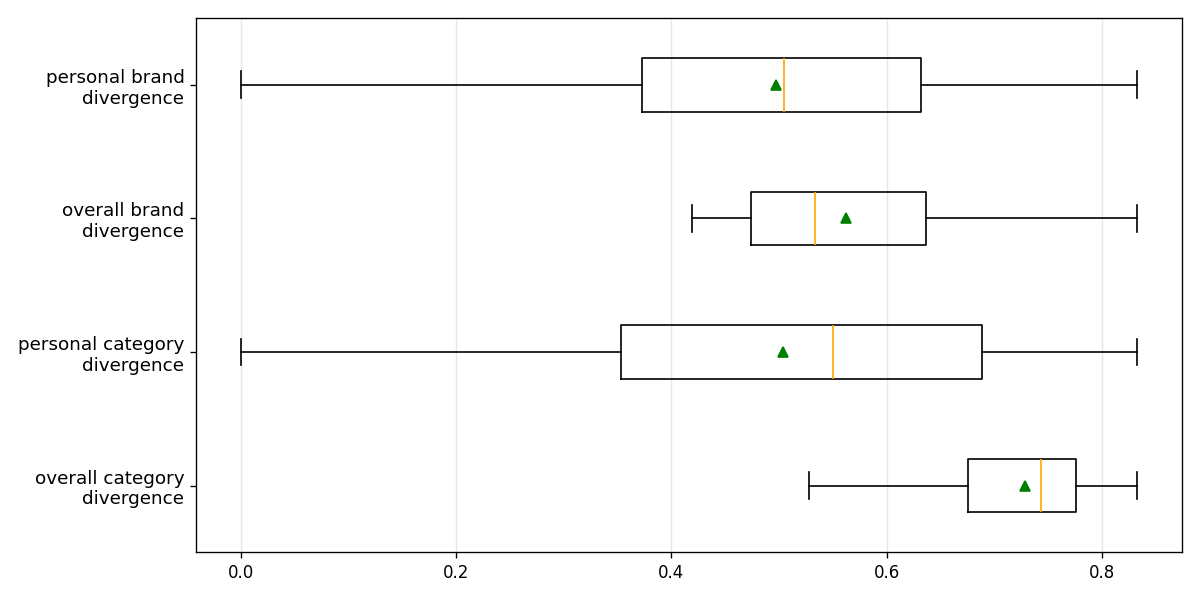}
    \caption{The boxplot of personal and overall interest
    divergence across users.}
    \label{fig:interest_divergence}
\end{wrapfigure}
To assess whether behavioral histories in KuaiSearch encode user-specific preference signals, we compare each user's recent interests against two references: their own earlier behavior and the aggregated earlier behavior of all eligible users. For users with more than ten interactions, the latest ten clicked or purchased items represent recent interests, while the remaining interactions characterize long-term preferences. We model interest distributions at the brand and first-level category levels and quantify their differences using Jensen--Shannon (JS) divergence. As shown in Figure~\ref{fig:interest_divergence}, recent interests are consistently closer to users' own histories than to the population distribution, with a larger gap at the category level than at the brand level. The broad divergence distributions further indicate temporal variation in user preferences. These results demonstrate that KuaiSearch captures both stable personalized interests and evolving behavioral patterns.

\subsubsection{Query Analysis.}

\begin{wrapfigure}{r}{0.5\textwidth}
    \centering
    \includegraphics[width=\linewidth]
    {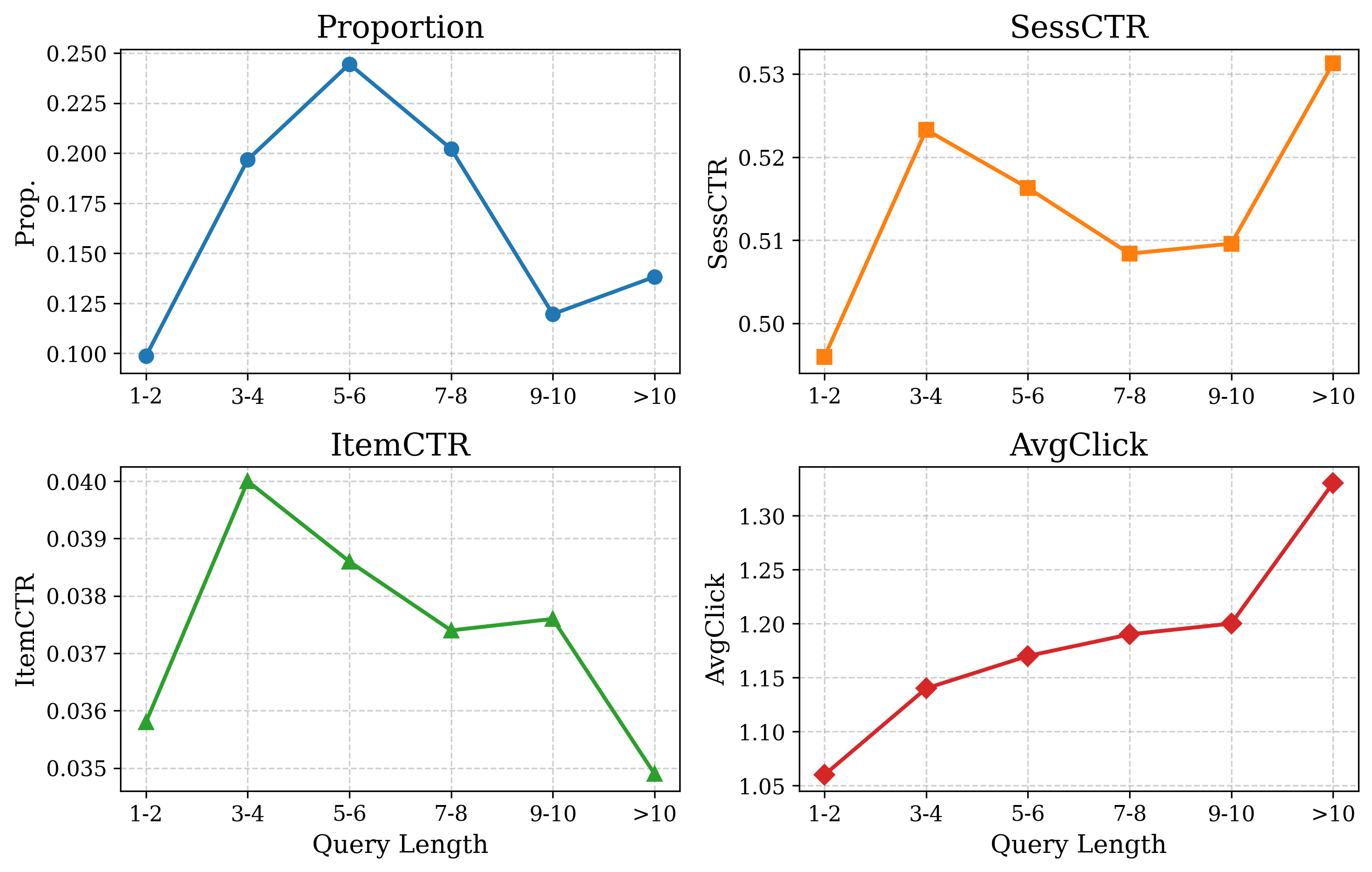}
    \caption{Engagement statistics across different query lengths.}
    \label{fig:query_length_engagement}
\end{wrapfigure}

In this subsection, we provide an analysis of the search queries in KuaiSearch, focusing on query length distributions, user interaction patterns, and query intent ambiguity.

Figure~\ref{fig:query_length_engagement} summarizes user engagement across different query length ranges. Medium-length queries dominate the distribution, with lengths of 5--6 and 7--8 together accounting for 44.68\% of all requests. The session-level click-through rate (SessCTR) ranges from 0.496 to 0.531 across query lengths and does not exhibit a clear monotonic trend; nevertheless, very long queries ($>10$) attain the highest SessCTR, suggesting a higher likelihood of triggering at least one click under more specific intent expressions. In contrast, the average number of clicked items per session (AvgClick) increases steadily with query length, rising from 1.06 to 1.33. This trend suggests that longer queries tend to express more specific user intents and lead to deeper interaction with the ranked results. Item-level click-through rate (ItemCTR) peaks at short-to-medium query lengths (3--4) and then gradually declines as query length increases, with a noticeable drop for very long queries ($>10$). One possible explanation is that stronger intent encourages broader browsing and comparison behavior, dispersing clicks across a larger set of exposed items rather than concentrating them on a few.

\begin{wrapfigure}{r}{0.50\textwidth}
    \vspace{-10pt}
    \centering
    \includegraphics[width=\linewidth]
        {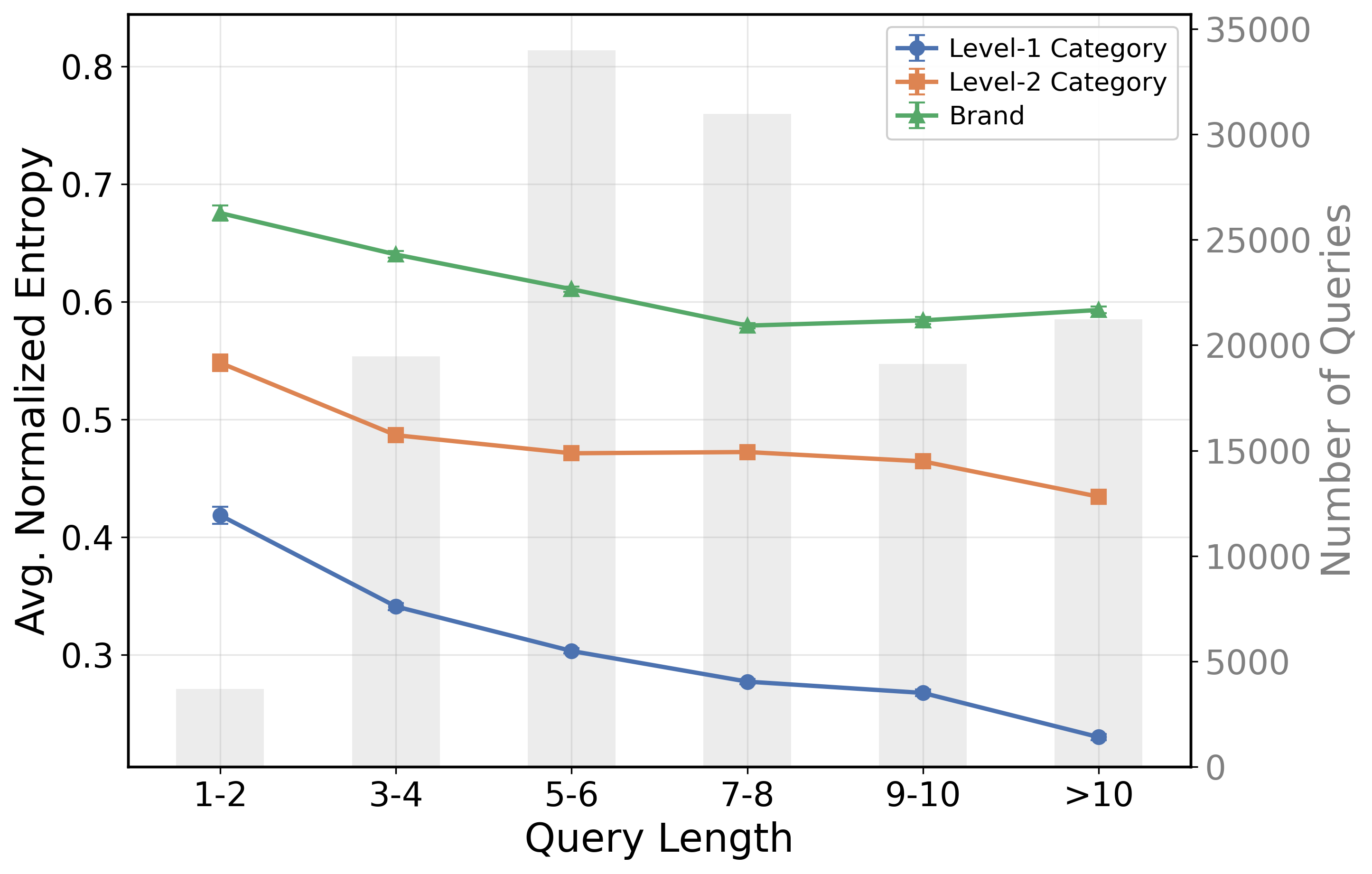}
    \caption{Query-level intent ambiguity across query lengths.}
    \label{fig:query_ambiguity_combined}
    \vspace{-8pt}
\end{wrapfigure}
Beyond engagement patterns, we further examine query-level intent ambiguity through the category and brand distributions of positively interacted items, including both clicks and purchases. For each query, we aggregate all clicked and purchased products across sessions and deduplicate them to construct a unique interacted-item set. We then compute the normalized entropy of the first-level category, second-level category, and brand distributions within this set, where a higher entropy indicates that the query is associated with a more diverse range of potential product intents. As shown in Figure~\ref{fig:query_ambiguity_combined}, category-level ambiguity generally decreases as query length increases. Specifically, the average first-level category entropy declines from 0.419 for queries of 1--2 characters to 0.230 for queries longer than 10 characters, while the corresponding second-level category entropy decreases from 0.548 to 0.434. These findings suggest that longer queries typically provide more explicit constraints, thereby narrowing the range of potentially relevant product categories. In contrast, brand-level entropy remains consistently high across all query-length groups, ranging from approximately 0.58 to 0.68. This pattern indicates that even relatively detailed queries often convey limited information about users' brand preferences. Overall, short queries exhibit particularly pronounced category-level ambiguity, whereas fine-grained brand preferences may require additional personalized signals derived from users' long-term historical interaction behaviors.

\subsubsection{Vocabulary Asymmetry between Queries and Item Titles}

\begin{figure}[t]
    \centering

    \begin{minipage}[t]{0.35\linewidth}
        \centering
        \includegraphics[width=\linewidth]
            {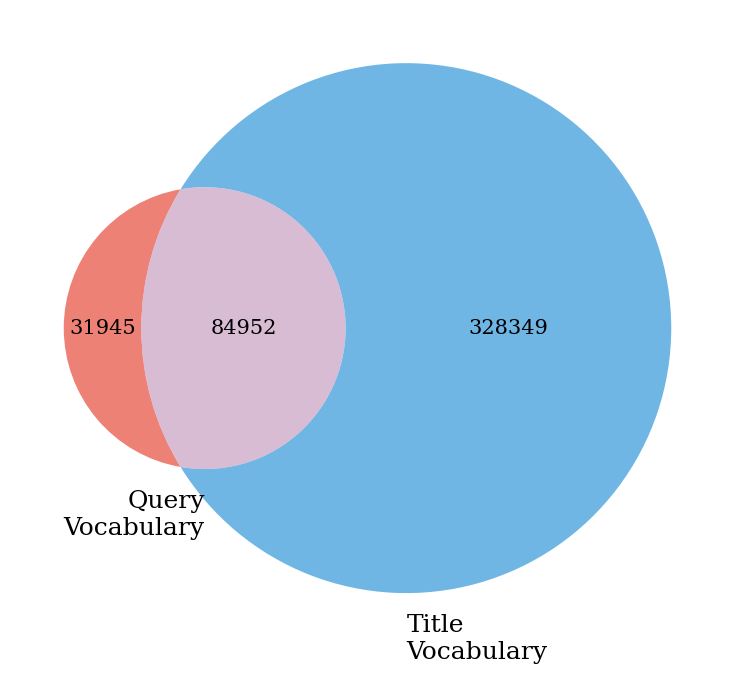}
        \\[-1mm]
        {\small (a) Vocabulary Overlap}
    \end{minipage}
    \hspace{0.05\linewidth}
    \begin{minipage}[t]{0.38\linewidth}
        \centering
        \includegraphics[width=\linewidth]
            {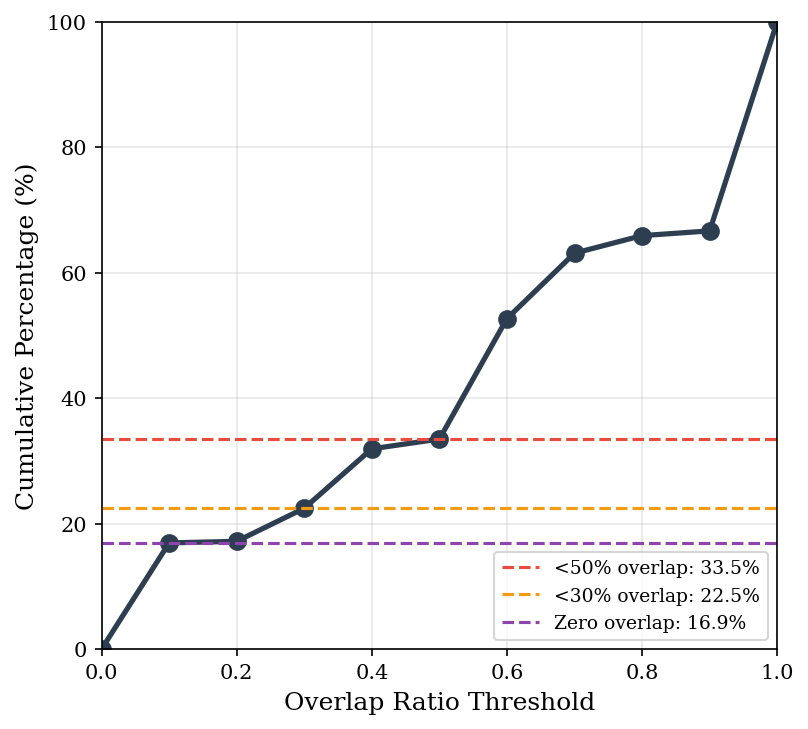}
        \\[-1mm]
        {\small (b) Cumulative Distribution}
    \end{minipage}

    \caption{Vocabulary asymmetry and token overlap between user
    queries and item titles.}
    \label{fig:vocabulary}
\end{figure}

This subsection analyzes the lexical gap between user queries and item titles in KuaiSearch. We use 3,035,445 clicked query--title pairs, tokenize texts with Jieba\footnote{\url{https://github.com/fxsjy/jieba}}, and define token overlap as the proportion of query tokens appearing in the clicked item title. As shown in Figure~\ref{fig:vocabulary}, item titles contain 413,301 unique tokens, 3.54 times the query vocabulary of 116,897. Among them, 328,349 tokens appear only in titles, 10.3 times the 31,945 query-exclusive tokens. This asymmetry reflects different expression patterns: queries are concise and focus on categories or key attributes, while titles contain richer product information. At the instance level, 16.92\% of clicked pairs have no token overlap and 33.48\% have overlap below 50\%. These results show that exact term matching alone is insufficient for many positive pairs. Thus, lexical methods such as BM25~\citep{bm25} may struggle with synonymy, implicit attributes, and cross-granularity matching, making KuaiSearch a challenging benchmark for semantic retrieval.

\subsection{Relevance Data Analysis}
\label{sec:relevance}

\begin{wraptable}{R}{0.5\textwidth}
    \centering
    \renewcommand{\arraystretch}{1.08}
    \setlength{\tabcolsep}{3pt}

    \caption{Distribution of relevance scores.}
    \label{tab:score_distribution}

    \begin{tabular}{@{}lrrrr@{}}
        \toprule
        \textbf{Score}
        & \textbf{0}
        & \textbf{1}
        & \textbf{2}
        & \textbf{3} \\
        \midrule
        Number
        & 6,095
        & 8,834
        & 15,774
        & 15,719 \\
        Proportion
        & 13.13\%
        & 19.03\%
        & 33.98\%
        & 33.86\% \\
        \bottomrule
    \end{tabular}
\end{wraptable}
The relevance data in KuaiSearch are annotated by e-commerce experts using a four-level graded relevance scheme. A score of 0 denotes clearly irrelevant items, typically due to category or attribute mismatches; 1 indicates weak relevance, where items belong to related categories but fail to satisfy the search intent; 2 represents partially relevant items that match the general category but violate important constraints such as brand or key attributes; and 3 indicates highly relevant items that fully satisfy the user’s intent. This scheme provides fine-grained supervision for relevance modeling. Representative examples are shown in Table~\ref{tab:relevance_cases}. As shown in Table~\ref{tab:score_distribution}, scores 2 and 3 account for most instances, indicating that KuaiSearch contains many non-trivial partially and highly relevant pairs. These cases are difficult to distinguish because they require reasoning over subtle category boundaries and attribute-level constraints. Overall, KuaiSearch provides a challenging benchmark for fine-grained relevance modeling in realistic e-commerce search.







\begin{table}[t]
    \centering
    \caption{KuaiSearch-Lite recall results; best and second-best values are bold and underlined.}
    \label{tab:recall_results}

    \resizebox{\linewidth}{!}{
    \begin{tabular}{lccccccccc}
        \toprule
        & \multicolumn{2}{c}{\textit{Lexical matching}}
        & \multicolumn{4}{c}{\textit{Embedding-based}}
        & \multicolumn{3}{c}{\textit{Generative retrieval}} \\
        \cmidrule(lr){2-3}
        \cmidrule(lr){4-7}
        \cmidrule(lr){8-10}

        \textbf{Metric}
        & \textbf{BM25}
        & \textbf{DocT5Query}
        & \textbf{DPR-SDE}
        & \textbf{DPR-ADE}
        & \textbf{BGE-base}
        & \shortstack{\textbf{BGE-base}\\\textbf{(fine-tuned)}}
        & \textbf{DSI}
        & \textbf{LTRGR}
        & \textbf{LC-Rec} \\
        \midrule

        R@10
        & 0.071
        & 0.078
        & \underline{0.083}
        & 0.082
        & 0.074
        & \textbf{0.097}
        & 0.062
        & 0.069
        & 0.071 \\

        HR@10
        & 0.100
        & 0.110
        & \underline{0.121}
        & 0.118
        & 0.107
        & \textbf{0.138}
        & 0.097
        & 0.105
        & 0.109 \\

        R@50
        & 0.156
        & 0.177
        & \underline{0.208}
        & 0.203
        & 0.176
        & \textbf{0.229}
        & 0.137
        & 0.150
        & 0.148 \\

        HR@50
        & 0.209
        & 0.238
        & \underline{0.277}
        & 0.271
        & 0.235
        & \textbf{0.303}
        & 0.202
        & 0.218
        & 0.218 \\

        \bottomrule
    \end{tabular}
    }
\end{table}








\section{Experiment}

In this section, we report the performance of baseline approaches on recall, ranking, and relevance tasks. All experiments are conducted on our KuaiSearch-Lite benchmark, which improves experimental efficiency while preserving the representativeness of the evaluation. For trainable models, we report the average results over three runs using different seeds.

\subsection{Recall Task Performance}

In the recall task, we retain search requests with at least one click or purchase, using the last day of KuaiSearch-Lite recall logs for testing and the remaining logs for training. We evaluate Recall@K (R@K) and Hit Rate@K (HR@K), where R@K measures the proportion of interacted items retrieved in the top-$K$, and HR@K indicates whether any interacted item is retrieved. Baselines include lexical methods (BM25, DocT5Query \citep{bm25,doc2query}), embedding-based methods (DPR-ADE, DPR-SDE, BGE-base-zh, and BGE-base-zh (fine-tuning) \citep{dpr,adesde,bge}), and generative methods (DSI, LTRGR, and LC-Rec \citep{dsi,ltrgr,lcrec}). DPR-ADE uses separate query and item encoders, while DPR-SDE and both BGE variants use shared encoders; BGE-base-zh (fine-tuning) is adapted on the training set. All methods use item titles. For DocT5Query, we train an mT5-base model \citep{mt5} on the training set to generate 10 pseudo-queries per item, concatenated with item titles as indexed text. For generative retrieval, we use RQ-Kmeans \citep{qarm,onerec} to assign semantic IDs with a three-layer codebook of size 512 per layer. For items sharing the same semantic ID, an additional deduplication position is appended to ensure uniqueness. Further implementation details are provided in Appendix~\ref{app:details}.

Table~\ref{tab:recall_results} reports the performance of different retrieval methods in terms of Recall and Hit Rate. Embedding-based retrieval methods outperform both lexical matching and generative retrieval methods across most evaluation metrics. Notably, BGE-base-zh after fine-tuning achieves the best performance across all metrics, demonstrating the effectiveness of adapting pretrained text representations to the e-commerce search domain. DPR-SDE also outperforms DPR-ADE, indicating its superior ability to model semantic relevance between queries and items. This observation is consistent with previous studies \citep{adesde}, which show that symmetric dual-encoder architectures with shared parameters can alleviate misalignment in the embedding space, improving retrieval performance. In contrast, generative retrieval methods perform worse than embedding-based methods and achieve performance comparable to lexical matching methods. This gap mainly arises from the inherent complexity of e-commerce search: user queries are often ambiguous and multi-intent, product catalogs are vast with a long-tail distribution, and product texts are more noisy and unstructured compared with traditional retrieval settings. These factors make it particularly challenging for generative models to generate precise item identifiers during the recall phase.

\subsection{Ranking Task Performance}

In the ranking task, we use data from the last day of the KuaiSearch-Lite ranking dataset as the test set, while all preceding data are used for training. This chronological split reflects practical deployment scenarios, in which models are trained on historical interactions and evaluated on future user behavior. We evaluate model performance using ROC-AUC, which measures the ability to distinguish clicked items from non-clicked ones. We consider eight representative CTR prediction models: DNN \citep{dnn}, Wide \& Deep \citep{widedeep}, DCN \citep{dcn1}, DCN-v2 \citep{dcn2}, DIN \citep{din}, GDCN \citep{gdcn}, WuKong \citep{wukong}, and FinalMLP \citep{finalmlp}. These baselines cover diverse architectural paradigms, including multilayer perceptron-based models, explicit feature-crossing networks, attention-based user behavior modeling, and methods designed to capture high-order feature interactions. To ensure a fair comparison, all models use the same feature set, including user demographics such as gender and age, the 20 most recent clicked items and their categories, the target item and its category, and query and item-title embeddings extracted using a pretrained BGE-small-zh-v1.5 encoder \citep{bge}. The data split and preprocessing procedures are kept consistent across all baselines. Further implementation details are provided in Appendix~\ref{app:details}.

\begin{table}[t]
    \centering
    \renewcommand{\arraystretch}{1.12}
    \caption{Ranking ROC-AUC; best and second-best values are bold and underlined.}
    \label{tab:ctr}

    \resizebox{\linewidth}{!}{
    \begin{tabular}{lcccccccc}
        \toprule
        \textbf{Metric}
        & \textbf{DNN}
        & \textbf{Wide \& Deep}
        & \textbf{DCN}
        & \textbf{DCN-v2}
        & \textbf{DIN}
        & \textbf{GDCN}
        & \textbf{WuKong}
        & \textbf{FinalMLP} \\
        \midrule

        ROC-AUC
        & \underline{0.6258}
        & 0.6217
        & 0.6194
        & 0.6239
        & \textbf{0.6262}
        & 0.6224
        & 0.6243
        & 0.6220 \\

        \bottomrule
    \end{tabular}
    }
\end{table}

Table~\ref{tab:ctr} reports the ROC-AUC performance of the evaluated ranking models. The results lie within a narrow range of 0.6194--0.6262, indicating broadly comparable performance under identical feature and training settings. Nevertheless, in large-scale e-commerce systems, even modest offline improvements may translate into meaningful online gains. DIN achieves the best ROC-AUC of 0.6262, suggesting that attention-based interest modeling effectively identifies relevant historical behaviors. DNN follows closely with 0.6258, showing that a simple multilayer perceptron remains competitive when rich user, item, behavioral, and textual features are available. Among feature-interaction models, DCN-v2 improves over DCN from 0.6194 to 0.6239, while WuKong reaches 0.6243 through explicit high-order interaction modeling. However, GDCN and FinalMLP do not outperform DNN or DIN, indicating that under the current experimental setting, more complex interaction modules do not necessarily yield better performance. Overall, the results highlight the importance of both behavioral preference modeling and input features in e-commerce search ranking.

\subsection{Relevance Task Performance}

In the relevance task, we split the dataset, using 10\% of samples for testing and the rest for training. Samples annotated with a relevance score of 3 are treated as positive instances, while all others are regarded as negative, formulating relevance modeling as a binary classification problem. Model performance is evaluated via ROC-AUC and PR-AUC: ROC-AUC measures ranking quality, and PR-AUC better reflects the ability to identify positive instances under class-imbalanced settings.

We evaluate three representative and widely-used relevance modeling approaches: bi-encoder-based methods (BGE \citep{bge}), cross-encoder-based methods (BERT-Chinese, BERT-Multilingual \citep{bert}, and XLM-RoBERTa \citep{xlm}), and LLM-based generative classification methods (Qwen3 \citep{qwen3} and Llama3.2 \citep{llama3}). For the LLM-based generative classification approach, each query-item pair is represented as a natural language input along with the task prompt, and the candidate labels (``yes'' and ``no'') are treated as target tokens for LoRA fine-tuning \citep{lora}. During inference, the model performs a single forward pass without autoregressive decoding, and the conditional probability distribution over the candidate label tokens is extracted at the final token position. The probabilities corresponding to ``yes'' and ``no'' are normalized via a softmax operation, and the probability of ``yes'' is used as the final relevance score for subsequent ranking and binary classification evaluation. Further implementation details are provided in Appendix~\ref{app:details}.

As shown in Table~\ref{tab:embedding_auc}, bi-encoder-based methods exhibit weaker performance. This is primarily due to the fact that queries and items are encoded independently and matched via vector similarity, limiting the model’s ability to capture fine-grained semantic interactions required for accurate relevance judgment. In contrast, cross-encoder-based methods achieve better performance, validating the effectiveness of explicitly modeling token-level interactions between queries and items. The performance of generative classification models varies across model families and scales. Among all evaluated models, Qwen3-1.7B achieves the best performance. In particular, Qwen3-1.7B achieves the highest scores on both ROC-AUC and PR-AUC, substantially outperforming the bi-encoder and cross-encoder baselines. Notably, Qwen3-0.6B surpasses Llama3.2-1B and achieves performance comparable to Llama3.2-3B, demonstrating the strong modeling capability and favorable parameter efficiency of the Qwen3 family for relevance modeling. Overall, these results suggest that models capable of capturing fine-grained semantic relationships between queries and items perform better in relevance modeling, and generative classification methods exhibit significant potential. A complementary ablation analysis on product metadata is provided in Appendix~\ref{sec:ablation_relevance}.

\begin{table}[t]
    \centering
    \footnotesize
    \renewcommand{\arraystretch}{1.15}
    \setlength{\tabcolsep}{3pt}

    \caption{Relevance results; best and second-best values are bold and underlined.}
    \label{tab:embedding_auc}

    \begin{tabular*}{\linewidth}{
        @{\extracolsep{\fill}}lcccccccccc@{}
    }
        \toprule

        & \multicolumn{2}{c}{\textsc{Bi-encoder}}
        & \multicolumn{4}{c}{\textsc{Cross-encoder}}
        & \multicolumn{4}{c}{\textsc{Generative}} \\
        \cmidrule(lr){2-3}
        \cmidrule(lr){4-7}
        \cmidrule(lr){8-11}

        \textbf{Metric}
        & \shortstack{\textbf{BGE}\\\textbf{Base}}
        & \shortstack{\textbf{BGE}\\\textbf{Large}}
        & \shortstack{\textbf{BERT}\\\textbf{Chinese}}
        & \shortstack{\textbf{BERT}\\\textbf{Multi.}}
        & \shortstack{\textbf{XLM-R}\\\textbf{Base}}
        & \shortstack{\textbf{XLM-R}\\\textbf{Large}}
        & \shortstack{\textbf{Qwen3}\\\textbf{0.6B}}
        & \shortstack{\textbf{Qwen3}\\\textbf{1.7B}}
        & \shortstack{\textbf{Llama3.2}\\\textbf{1B}}
        & \shortstack{\textbf{Llama3.2}\\\textbf{3B}} \\
        \midrule

        Size
        & 0.10B & 0.32B
        & 0.11B & 0.11B
        & 0.27B & 0.55B
        & 0.60B & 1.70B
        & 1.00B & 3.00B \\
        \midrule

        ROC-AUC
        & 0.7475 & 0.7531
        & 0.7606 & 0.7737
        & 0.7941 & 0.8005
        & 0.7994 & \textbf{0.8215}
        & 0.7602 & \underline{0.8093} \\

        PR-AUC
        & 0.5791 & 0.6052
        & 0.6041 & 0.6383
        & 0.6658 & \underline{0.6756}
        & 0.6524 & \textbf{0.6966}
        & 0.5927 & 0.6696 \\

        \bottomrule
    \end{tabular*}
\end{table}
\section{Conclusion}
We introduced KuaiSearch, a large-scale e-commerce search dataset constructed from authentic Kuaishou interactions. It preserves real queries, natural-language product texts, structured metadata, and diverse user behaviors without popularity-based filtering, while retaining long-tail products and cold-start users. KuaiSearch supports three key e-commerce search tasks: \emph{recall}, \emph{ranking}, and \emph{relevance judgment}. Our analyses highlight query ambiguity, lexical gaps between queries and product titles, and heterogeneous user preferences, revealing practical challenges in real-world e-commerce search. Benchmark experiments further demonstrate its value for studying personalized retrieval, ranking, and fine-grained semantic relevance, while providing reference results for representative methods. We hope KuaiSearch serves as a useful resource for developing and evaluating more robust LLM-based e-commerce search systems.

\section{Limitations}
KuaiSearch reflects a Chinese e-commerce platform and contains linguistic, cultural, and domain-specific biases associated with its language, user population, product catalog, and shopping practices. Chinese expressions and tokenization may affect benchmark results, which should not be generalized to other languages or markets without validation. Translation may alter query ambiguity and does not reproduce authentic shopping behavior in the target market. Even after translation, differences in product availability, brand familiarity, and shopping conventions may limit the transferability of the observed findings. Future work should evaluate models using datasets from diverse languages, markets, and user populations to assess cross-domain generalizability and identify findings that remain consistent across settings.

\section*{AI Use Statement}
Generative AI tools were used to assist with language editing and literature retrieval and discovery. Specifically, they were used to improve grammar, clarity, and phrasing, and to help identify potentially relevant prior work. All AI-assisted edits and literature suggestions were reviewed and verified by the authors. Generative AI tools were not used for research ideation or execution, drafting substantive sections of the paper, generating synthetic datasets, or proving mathematical claims. The authors take full responsibility for the final content of this work.

\section*{Ethics Statement}
The construction and release of KuaiSearch follow Kuaishou Technology's
data privacy and security governance procedures. All released data are
de-identified, with identifiers remapped, demographic attributes
aggregated, geographic information coarsened, and absolute timestamps
converted to relative indices. The data release underwent internal legal
and data-security compliance review. Further details are provided in
Appendix~\ref{app:privacy}.

\bibliography{iclr2027_conference}
\bibliographystyle{iclr2027_conference}

\appendix

\section{Relevance Annotation Quality and Reliability}
\label{sec:annotation_quality}

The quality and reliability of the relevance annotations are fundamental to the validity of the model evaluation results reported in this work. To ensure the accuracy and consistency of the annotated data, we designed and implemented a rigorous annotation procedure and quality control mechanism. All query--item pairs were independently annotated by three annotators, who had no access to the decisions made by the other annotators. The annotators had frontline industrial experience in e-commerce search algorithms and received comprehensive and systematic training based on predefined annotation guidelines before the formal annotation process. In addition, annotators were required to provide an explicit written justification for each assigned label to support subsequent quality control. This mandatory requirement encouraged annotators to make decisions according to predefined criteria rather than arbitrary subjective judgments. Before applying majority voting, we first measured inter-annotator agreement on the raw annotations. The results show that the three annotators reached agreement on 93\% of the relevance labels, with a Fleiss' $\kappa$ score of 0.89. For samples with annotation disagreements, we adopted a tiered resolution strategy. General disagreements were resolved through majority voting, whereas samples with substantial disagreements first underwent structured group discussions, during which the annotators re-examined the annotation guidelines and specific cases, followed by a second round of independent re-annotation. These annotation procedures and quality control measures effectively reduced inconsistency and arbitrariness in the annotations, providing a reliable foundation for evaluating relevance models.

\section{Data Privacy and Ethical Compliance}
\label{app:privacy}
The development and release of the KuaiSearch dataset follow Kuaishou Technology's data privacy governance framework. All data collection, processing, and release procedures underwent internal legal and data security compliance reviews and were conducted in accordance with the relevant requirements of the Personal Information Protection Law of the People's Republic of China. The compliance review further confirmed that the relevant user authorization is covered by the applicable platform privacy policy. Before public release, we applied multiple de-identification measures to protect user privacy. Specifically, all user, item, seller, and other related identifiers were replaced with irreversible random identifiers. Age information was aggregated into seven predefined age groups, geographic information was retained only at the provincial level for users, and all absolute timestamps were converted into relative time indices. The released dataset contains no direct personal identifiers and cannot be linked back to original platform accounts. These de-identification procedures, together with the internal compliance review process, substantially reduce potential re-identification risks. 

\section{Construction of KuaiSearch-Lite}
\label{app:lite_construction}

To facilitate efficient experimentation while preserving realistic user behavior patterns, we construct KuaiSearch-Lite by randomly selecting 102,086 users from the full KuaiSearch dataset and retaining their complete search and interaction records within a continuous one-month period. The subset is constructed at the user level rather than by independently sampling individual search requests or interaction instances. This design preserves the temporal continuity of user behavior histories and maintains the associations among queries, exposed items, clicks, and purchases. For each selected user, we retain the corresponding user information, search requests, exposed items, clicked and purchased items, and associated product metadata to construct the recall and ranking data. Consequently, KuaiSearch-Lite preserves the main behavioral and structural characteristics of the full dataset while substantially reducing the computational cost of model training and evaluation. The resulting subset contains 555,553 queries and 6,634,118 products, providing an efficient and representative benchmark for model development, ablation studies, and reproducible evaluation.

\section{Experimental Details}
\label{app:details}

This section reports the key implementation details of the recall, ranking, and relevance experiments. More comprehensive model-specific configurations, preprocessing procedures, and training and evaluation scripts are available in our released codebase.
\subsection{Recall Experimental Details}

For lexical retrieval, BM25 is implemented with Jieba-based Chinese word segmentation, and the hyperparameters are set to $k_1=1.2$ and $b=0.75$. For dense retrieval methods, the maximum sequence lengths of both queries and item titles are set to 64 tokens. Query--item relevance is measured by inner-product similarity, and all item representations are indexed using \texttt{Faiss IndexFlatIP}, enabling exact maximum-inner-product search over the candidate corpus. For generative retrieval methods, we adopt prefix-tree-constrained decoding to restrict the generation space to valid semantic item identifiers. DSI and LTRGR are implemented with mT5-base as the backbone, whereas LC-Rec adopts Qwen3-0.6B-Instruct.

\subsection{Ranking Experimental Details}

All ranking baselines use the same feature and training configurations. The embedding dimensions of user IDs and item IDs are both set to 32, while those of gender, age group, and category levels 1--3 are set to 8, 16, 16, 32, and 64, respectively. User behavior sequences retain the 20 most recent clicked items. All models are optimized using Adam with an initial learning rate of $1\times10^{-3}$, a weight decay of $1\times10^{-5}$, and binary cross-entropy loss. The batch size is 20,000, and the maximum number of training epochs is 20. 

\subsection{Relevance Experimental Details}

For all relevance models, the maximum input sequence length is set to 512 tokens; for bi-encoders, this limit is applied separately to the query and item inputs. Bi-encoder models encode queries and items independently and compute relevance scores using cosine similarity between their representations. Cross-encoder models concatenate and jointly encode query and item texts to capture fine-grained token-level interactions. Qwen3-0.6B, Qwen3-1.7B, Llama3.2-1B, and Llama3.2-3B are fine-tuned using LoRA with rank $r=8$, scaling factor $\alpha=32$, and dropout rate 0.1. LoRA adapters are applied to the major linear projection modules, including \texttt{q\_proj}, \texttt{k\_proj}, \texttt{v\_proj}, \texttt{o\_proj}, \texttt{gate\_proj}, \texttt{up\_proj}, and \texttt{down\_proj}.

\section{Ablation Studies}
\label{sec:ablation}
\subsection{Ranking}
\label{sec:ablation_ranking}

We conduct ablation experiments on feature groups for the ranking task using the best-performing DIN model to quantify the contribution of different signal types. Table~\ref{tab:ablation_ranking} reports the ROC-AUC performance when removing each feature group individually while keeping all others intact.

\begin{table}[htbp]
    \centering
    \caption{Ablation study on feature groups for ranking (DIN).}
    \label{tab:ablation_ranking}
    \begin{tabular}{lc}
        \toprule
        \textbf{Setting} & \textbf{ROC-AUC} \\
        \midrule
        DIN       & 0.6262 \\
        w/o text         & 0.6078 \\
        w/o history      & 0.6206 \\
        w/o demographics & 0.6197 \\
        w/o category     & 0.6249 \\
        \bottomrule
    \end{tabular}
\end{table}

The results show that textual features provide the most important signal for ranking, as their removal causes the largest ROC-AUC drop from 0.6262 to 0.6078. This finding highlights the central role of query--item semantic matching in e-commerce ranking, as textual representations directly capture the alignment between user queries and product descriptions. Demographic and historical behavior features also contribute substantially, with their removal reducing ROC-AUC to 0.6197 and 0.6206, respectively. These results suggest that both group-level user characteristics and sequential behavioral preferences provide complementary signals for personalized ranking. Category features yield a smaller but still positive contribution, with ROC-AUC decreasing to 0.6249 when they are removed. Overall, these findings demonstrate the value of the heterogeneous multi-field features provided by KuaiSearch for developing and evaluating ranking models.

\subsection{Relevance}
\label{sec:ablation_relevance}

We conduct ablation experiments on the product metadata inputs for the relevance task using the parameter-efficient Qwen3-0.6B model. Table~\ref{tab:ablation_relevance} reports the PR-AUC performance when progressively adding different metadata fields, starting from the basic product title (T) and incrementally including brand (B), seller name (S), and attribute (A) information.

\begin{table}[htbp]
    \centering
    \small
    \caption{Ablation study on product metadata for relevance (Qwen3-0.6B).}
    \label{tab:ablation_relevance}
    \begin{tabular}{lcccc}
        \toprule
        \textbf{Setting} & T & T+B & T+B+S & T+B+S+A \\
        \midrule
        PR-AUC & 0.6383 & 0.6414 & 0.6432 & 0.6524 \\
        \bottomrule
    \end{tabular}
\end{table}

The results show that relevance performance consistently improves as richer product metadata is introduced, with attribute information bringing the largest incremental gain of 0.0092 PR-AUC. This is because attribute fields contain fine-grained product specifications such as size, material, style, and functional features, which are critical for distinguishing between products that share similar titles but differ in key characteristics that directly match user intent. 

\section{Long-tail Scenario Analysis}
\label{sec:longtail_analysis}

We further conduct a long-tail analysis on the ranking task using the best-performing DIN to evaluate performance across different popularity segments. We define head items as the top 20\% most frequently interacted products, while the remaining 80\% of products are considered tail items. Similarly, head users are defined as the top 20\% most active users with the most search sessions, while the remaining users are regarded as tail users.

\begin{table}[htbp]
    \centering
    \caption{DIN model performance on head and tail segments}
    \label{tab:longtail_performance}
    \begin{tabular}{lcc}
        \toprule
        \textbf{Group} & \textbf{Head} & \textbf{Tail} \\
        \midrule
        Item ROC-AUC & 0.6606 & 0.6180 \\
        User ROC-AUC & 0.6473 & 0.6189 \\
        \bottomrule
    \end{tabular}
\end{table}

These results reveal a substantial performance gap between head and tail segments. The model achieves substantially higher ROC-AUC on head items and head users, which can be attributed to the abundance of interaction data available for these segments, enabling more reliable statistical feature estimation and more accurate preference modeling. In contrast, tail items suffer from severe data sparsity and noisy statistical features, while tail users have limited historical behavior signals, making personalized ranking particularly challenging. These findings highlight that long-tail ranking remains a critical challenge in e-commerce search.

\begin{CJK}{UTF8}{gbsn}
\begin{table*}[t]
\centering
\small
\caption{Representative query--item pairs across relevance scores in KuaiSearch. Scores range from 0 (irrelevant) to 3 (highly relevant). }
\label{tab:relevance_cases}

\setlength{\tabcolsep}{3pt}
\renewcommand{\arraystretch}{1.08}

\begin{tabularx}{\textwidth}{
    c|
    >{\raggedright\arraybackslash}p{2.1cm}|
    >{\raggedright\arraybackslash}X|
    >{\raggedright\arraybackslash}p{1.6cm}|
    >{\raggedright\arraybackslash}p{2.5cm}|
    >{\raggedright\arraybackslash}p{3.0cm}
}
\toprule
\textbf{Score} &
\textbf{Query} &
\textbf{Item Title} &
\textbf{Brand} &
\textbf{Seller} &
\textbf{Key Attributes} \\
\midrule

3 &
眼影盘橘朵\newline
(Judydoll eyeshadow) &
Judydoll橘朵单色眼影哑光日常新手淡妆闪片牛郎\newline
(Judydoll single eyeshadow matte daily nude glitter) &
橘朵\newline
(Judydoll) &
俏俏美妆优选\newline
(Qiaoqiao Beauty Selection) &
易上色, 眼影盘, 珠光, 单色\newline
(easy color, eyeshadow palette, pearl, single color) \\
\midrule

2 &
帽子棉女童\newline
(Cotton girl's hat) &
爱贝迪拉婴儿帽子棉帽子四季款新生婴幼保暖胎帽护囟门帽宝宝可爱\newline
(Aibedila baby cotton hat, four seasons, keep warm, protect fontanel, cute) &
爱贝迪拉\newline
(Aibedila) &
爱贝迪拉母婴旗舰店\newline
(Aibedila Maternity \& Baby Flagship Store) &
胎帽, 棉, 翻边\newline
(fetal hat, cotton, turn-up edge) \\
\midrule

1 &
日奈绘画\newline
(Rina painting) &
节奏盒子描摹本画画本儿童线稿手绘描摹控笔练习绘画本图画本涂色\newline
(Rhythm box tracing book, children's line drawing, hand tracing, pen control practice, coloring book) &
艺丛\newline
(Yicong) &
彦和文具专营店\newline
(Yanhe Stationery Store) &
通用, 8岁--14岁, 艺丛\newline
(universal, ages 8--14, Yicong) \\
\midrule

0 &
防熊三件套\newline
(Bear-proof three-piece set) &
细细条冬天毛绒可爱小熊帽子围巾手套一体三件套骑车防寒护耳帽\newline
(Xixitiao winter plush cute bear hat scarf gloves set, cycling cold-proof earflap hat) &
细细条\newline
(Xixitiao) &
细细条旗舰店\newline
(Xixitiao Flagship Store) &
毛绒, 儿童, 细细条\newline
(plush, children, Xixitiao) \\
\bottomrule
\end{tabularx}
\end{table*}
\end{CJK}

\section{Data License and Intended Use} 
KuaiSearch is released under the Creative Commons Attribution 4.0 International (CC BY 4.0) license. It is intended to support research and development in e-commerce retrieval, ranking, relevance modeling, personalization, and related areas. Users should provide appropriate attribution and comply with applicable laws and regulations.

\section{Metric Definitions}
\label{sec:metric_definitions}

To ensure scientific reproducibility, we provide formal mathematical definitions for the engagement, retrieval and ranking, personalization, and query analysis metrics used in this paper.

\subsection{Engagement Metrics}

\begin{itemize}

\item \textbf{Session CTR (SessCTR)}: Proportion of search sessions with at least one click:
\[
\mathrm{SessCTR}
=
\frac{1}{|\mathcal{S}|}
\sum_{s\in\mathcal{S}}
\mathbb{I}\left(|\mathcal{C}_s|>0\right),
\]
where $\mathcal{S}$ is the set of search sessions, $\mathcal{C}_s$ is the clicked-item set for session $s$, and $\mathbb{I}(\cdot)$ is the indicator function.

\item \textbf{Item CTR}: Proportion of exposed items that are clicked:
\[
\mathrm{ItemCTR}
=
\frac{
\sum_{s\in\mathcal{S}}|\mathcal{C}_s|
}{
\sum_{s\in\mathcal{S}}|\mathcal{E}_s|
},
\]
where $\mathcal{E}_s$ is the exposed-item set for session $s$.

\item \textbf{AvgClick}: Average number of clicked items per search session:
\[
\mathrm{AvgClick}
=
\frac{1}{|\mathcal{S}|}
\sum_{s\in\mathcal{S}}|\mathcal{C}_s|.
\]

\end{itemize}

\subsection{Retrieval \& Ranking Metrics}

\begin{itemize}

\item \textbf{Recall@K (R@K)}: Measures the average proportion of positively interacted items retrieved in the top-$K$ results:
\[
\mathrm{R@K}
=
\frac{1}{|\mathcal{S}|}
\sum_{s\in\mathcal{S}}
\frac{
|\mathrm{TopK}(s)\cap \mathcal{I}_s|
}{
|\mathcal{I}_s|
},
\]
where $\mathcal{S}$ is the set of evaluated search requests, $\mathrm{TopK}(s)$ is the retrieved top-$K$ item set, and $\mathcal{I}_s$ is the set of clicked or purchased items for request $s$.

\item \textbf{Hit Rate@K (HR@K)}: Measures the proportion of search requests with at least one positively interacted item in the top-$K$ results:
\[
\mathrm{HR@K}
=
\frac{1}{|\mathcal{S}|}
\sum_{s\in\mathcal{S}}
\mathbb{I}
\left(
|\mathrm{TopK}(s)\cap \mathcal{I}_s|>0
\right),
\]
where $\mathbb{I}(\cdot)$ is the indicator function.

\item \textbf{ROC-AUC}: Measures the model's ability to distinguish positive from negative samples across classification thresholds.

\item \textbf{PR-AUC}: Measures performance across precision--recall trade-offs and is particularly informative under class imbalance.

\end{itemize}

\subsection{Personalization and Query Analysis Metrics}

\begin{itemize}

\item \textbf{Jensen--Shannon (JS) Divergence}: Measures the difference between two interest distributions $P$ and $Q$:
\[
\mathrm{JS}(P\|Q)
=
\frac{1}{2}D_{\mathrm{KL}}(P\|M)
+
\frac{1}{2}D_{\mathrm{KL}}(Q\|M),
\]
where $M=\frac{1}{2}(P+Q)$ and $D_{\mathrm{KL}}(\cdot\|\cdot)$ denotes the Kullback--Leibler divergence. A larger value indicates greater distributional discrepancy.

\item \textbf{Normalized Entropy}: Measures query-level intent ambiguity based on the attribute distribution of interacted items:
\[
\widetilde{H}(q)
=
-\frac{\sum_{i=1}^{N} p_i(q)\log p_i(q)}
{\log N},
\]
where $p_i(q)$ is the proportion of interacted items associated with the $i$-th attribute value, and $N$ is the number of values in the attribute space. A higher value indicates greater category or brand diversity and stronger intent ambiguity.

\item \textbf{Token Overlap Ratio}: Measures lexical overlap between a query and its corresponding item title:
\[
\mathrm{Overlap}(q,t)
=
\frac{|T_q \cap T_t|}
{|T_q|},
\]
where $T_q$ and $T_t$ denote the query and item-title token sets, respectively. A lower value indicates a larger lexical gap.

\end{itemize}

\end{document}

%% file: math_commands.tex
\usepackage{amsmath,amsfonts,bm}

\def\eqref#1{equation~\ref{#1}}

\def\1{\bm{1}}

\DeclareMathAlphabet{\mathsfit}{\encodingdefault}{\sfdefault}{m}{sl}
\SetMathAlphabet{\mathsfit}{bold}{\encodingdefault}{\sfdefault}{bx}{n}



%% file: iclr2027_conference.bib
@inproceedings{survey1,
  title={Challenges and research opportunities in ecommerce search and recommendations},
  author={Tsagkias, Manos and King, Tracy Holloway and Kallumadi, Surya and Murdock, Vanessa and De Rijke, Maarten},
  booktitle={ACM Sigir Forum},
  volume={54},
  number={1},
  pages={1--23},
  year={2021},
  organization={ACM New York, NY, USA}
}

@article{survey2,
  title={A brief survey of machine learning and deep learning techniques for e-commerce research},
  author={Zhang, Xue and Guo, Fusen and Chen, Tao and Pan, Lei and Beliakov, Gleb and Wu, Jianzhang},
  journal={Journal of Theoretical and Applied Electronic Commerce Research},
  volume={18},
  number={4},
  pages={2188--2216},
  year={2023},
  publisher={MDPI}
}

@inproceedings{queryvague1,
  title={Analysis of a very large web search engine query log},
  author={Silverstein, Craig and Marais, Hannes and Henzinger, Monika and Moricz, Michael},
  booktitle={Acm sigir forum},
  volume={33},
  number={1},
  pages={6--12},
  year={1999},
  organization={ACM New York, NY, USA}
}

@inproceedings{queryvague2,
  title={Quantifying query ambiguity with topic distributions},
  author={Yano, Yuki and Tagami, Yukihiro and Tajima, Akira},
  booktitle={Proceedings of the 25th ACM International on Conference on Information and Knowledge Management},
  pages={1877--1880},
  year={2016}
}

@inproceedings{titlevague1,
  title={A multi-task learning approach for improving product title compression with user search log data},
  author={Wang, Jingang and Tian, Junfeng and Qiu, Long and Li, Sheng and Lang, Jun and Si, Luo and Lan, Man},
  booktitle={Proceedings of the AAAI conference on artificial intelligence},
  volume={32},
  number={1},
  year={2018}
}

@inproceedings{amazonc4,
  title={Bridging Language and Items for Retrieval and Recommendation: Benchmarking LLMs as Semantic Encoders},
  author={Hou, Yupeng and Li, Jiacheng and Fu, Xiangjun and He, Zhankui and Yan, An and Chen, Xiusi and McAuley, Julian},
  booktitle={Proceedings of the 64th Annual Meeting of the Association for Computational Linguistics (Volume 1: Long Papers)},
  pages={3251--3265},
  year={2026}
}

@inproceedings{titlevague2,
  title={Generating e-commerce product titles and predicting their quality},
  author={De Souza, Jose GC and Kozielski, Michael and Mathur, Prashant and Chang, Ernie and Guerini, Marco and Negri, Matteo and Turchi, Marco and Matusov, Evgeny},
  booktitle={Proceedings of the 11th international conference on natural language generation},
  pages={233--243},
  year={2018}
}

@inproceedings{per1,
  title={A zero attention model for personalized product search},
  author={Ai, Qingyao and Hill, Daniel N and Vishwanathan, SVN and Croft, W Bruce},
  booktitle={Proceedings of the 28th ACM International Conference on Information and Knowledge Management},
  pages={379--388},
  year={2019}
}

@inproceedings{per2,
  title={Model-agnostic vs. model-intrinsic interpretability for explainable product search},
  author={Ai, Qingyao and Narayanan. R, Lakshmi},
  booktitle={Proceedings of the 30th ACM International Conference on Information \& Knowledge Management},
  pages={5--15},
  year={2021}
}

@inproceedings{per3,
  title={Learning a hierarchical embedding model for personalized product search},
  author={Ai, Qingyao and Zhang, Yongfeng and Bi, Keping and Chen, Xu and Croft, W Bruce},
  booktitle={Proceedings of the 40th International ACM SIGIR Conference on Research and Development in Information Retrieval},
  pages={645--654},
  year={2017}
}

@inproceedings{per4,
  title={A category-aware multi-interest model for personalized product search},
  author={Liu, Jiongnan and Dou, Zhicheng and Zhu, Qiannan and Wen, Ji-Rong},
  booktitle={Proceedings of the acm web conference 2022},
  pages={360--368},
  year={2022}
}

@article{onesearch,
  title={Onesearch: A preliminary exploration of the unified end-to-end generative framework for e-commerce search},
  author={Chen, Ben and Guo, Xian and Wang, Siyuan and Liang, Zihan and Lv, Yue and Ma, Yufei and Xiao, Xinlong and Xue, Bowen and Zhang, Xuxin and Yang, Ying and others},
  journal={arXiv preprint arXiv:2509.03236},
  year={2025}
}

@article{zhiding,
  title={Towards Context-aware Reasoning-enhanced Generative Searching in E-commerce},
  author={Liu, Zhiding and Chen, Ben and Cheng, Mingyue and Chen, Enhong and Li, Li and Lei, Chenyi and Ou, Wenwu and Li, Han and Gai, Kun},
  journal={arXiv preprint arXiv:2510.16925},
  year={2025}
}

@article{onevision,
  title={OneVision: An End-to-End Generative Framework for Multi-view E-commerce Vision Search},
  author={Zheng, Zexin and Dai, Huangyu and Mao, Lingtao and Sun, Xinyu and Liang, Zihan and Chen, Ben and Ding, Yuqing and Lei, Chenyi and Ou, Wenwu and Li, Han and others},
  journal={arXiv preprint arXiv:2510.05759},
  year={2025}
}

@inproceedings{amazonquery,
  title={Learning latent vector spaces for product search},
  author={Van Gysel, Christophe and de Rijke, Maarten and Kanoulas, Evangelos},
  booktitle={Proceedings of the 25th ACM international on conference on information and knowledge management},
  pages={165--174},
  year={2016}
}

@inproceedings{jdsearch,
  title={Jdsearch: A personalized product search dataset with real queries and full interactions},
  author={Liu, Jiongnan and Dou, Zhicheng and Tang, Guoyu and Xu, Sulong},
  booktitle={Proceedings of the 46th International ACM SIGIR Conference on Research and Development in Information Retrieval},
  pages={2945--2952},
  year={2023}
}

@inproceedings{qilin,
  title={Qilin: A multimodal information retrieval dataset with app-level user sessions},
  author={Chen, Jia and Dong, Qian and Li, Haitao and He, Xiaohui and Gao, Yan and Cao, Shaosheng and Wu, Yi and Yang, Ping and Xu, Chen and Hu, Yao and others},
  booktitle={Proceedings of the 48th International ACM SIGIR Conference on Research and Development in Information Retrieval},
  pages={3670--3680},
  year={2025}
}

@article{tf-idf,
  title={Term-weighting approaches in automatic text retrieval},
  author={Salton, Gerard and Buckley, Christopher},
  journal={Information processing \& management},
  volume={24},
  number={5},
  pages={513--523},
  year={1988},
  publisher={Elsevier}
}

@article{bm25,
  title={The probabilistic relevance framework: BM25 and beyond},
  author={Robertson, Stephen and Zaragoza, Hugo and others},
  journal={Foundations and trends{\textregistered} in information retrieval},
  volume={3},
  number={4},
  pages={333--389},
  year={2009},
  publisher={Now Publishers, Inc.}
}

@inproceedings{dpr,
  title={Dense Passage Retrieval for Open-Domain Question Answering.},
  author={Karpukhin, Vladimir and Oguz, Barlas and Min, Sewon and Lewis, Patrick SH and Wu, Ledell and Edunov, Sergey and Chen, Danqi and Yih, Wen-tau},
  booktitle={EMNLP (1)},
  pages={6769--6781},
  year={2020}
}

@article{ance,
  title={Approximate nearest neighbor negative contrastive learning for dense text retrieval},
  author={Xiong, Lee and Xiong, Chenyan and Li, Ye and Tang, Kwok-Fung and Liu, Jialin and Bennett, Paul and Ahmed, Junaid and Overwijk, Arnold},
  journal={arXiv preprint arXiv:2007.00808},
  year={2020}
}

@article{dsi,
  title={Transformer memory as a differentiable search index},
  author={Tay, Yi and Tran, Vinh and Dehghani, Mostafa and Ni, Jianmo and Bahri, Dara and Mehta, Harsh and Qin, Zhen and Hui, Kai and Zhao, Zhe and Gupta, Jai and others},
  journal={Advances in Neural Information Processing Systems},
  volume={35},
  pages={21831--21843},
  year={2022}
}

@inproceedings{ltrgr,
  title={Learning to rank in generative retrieval},
  author={Li, Yongqi and Yang, Nan and Wang, Liang and Wei, Furu and Li, Wenjie},
  booktitle={Proceedings of the AAAI Conference on Artificial Intelligence},
  volume={38},
  number={8},
  pages={8716--8723},
  year={2024}
}

@article{dsi-qg,
  title={Bridging the gap between indexing and retrieval for differentiable search index with query generation},
  author={Zhuang, Shengyao and Ren, Houxing and Shou, Linjun and Pei, Jian and Gong, Ming and Zuccon, Guido and Jiang, Daxin},
  journal={arXiv preprint arXiv:2206.10128},
  year={2022}
}

@article{onesug,
  title={OneSug: The Unified End-to-End Generative Framework for E-commerce Query Suggestion},
  author={Guo, Xian and Chen, Ben and Wang, Siyuan and Yang, Ying and Lei, Chenyi and Ding, Yuqing and Li, Han},
  journal={arXiv preprint arXiv:2506.06913},
  year={2025}
}

@article{msmarco,
  title={Ms marco: A human-generated machine reading comprehension dataset},
  author={Nguyen, Tri and Rosenberg, Mir and Song, Xia and Gao, Jianfeng and Tiwary, Saurabh and Majumder, Rangan and Deng, Li},
  year={2016}
}

@article{nq,
  title={Natural questions: a benchmark for question answering research},
  author={Kwiatkowski, Tom and Palomaki, Jennimaria and Redfield, Olivia and Collins, Michael and Parikh, Ankur and Alberti, Chris and Epstein, Danielle and Polosukhin, Illia and Devlin, Jacob and Lee, Kenton and others},
  journal={Transactions of the Association for Computational Linguistics},
  volume={7},
  pages={453--466},
  year={2019},
  publisher={MIT Press One Rogers Street, Cambridge, MA 02142-1209, USA journals-info~…}
}

@article{doc2query,
  title={Document expansion by query prediction},
  author={Nogueira, Rodrigo and Yang, Wei and Lin, Jimmy and Cho, Kyunghyun},
  journal={arXiv preprint arXiv:1904.08375},
  year={2019}
}

@article{onerec,
  title={Onerec: Unifying retrieve and rank with generative recommender and iterative preference alignment},
  author={Deng, Jiaxin and Wang, Shiyao and Cai, Kuo and Ren, Lejian and Hu, Qigen and Ding, Weifeng and Luo, Qiang and Zhou, Guorui},
  journal={arXiv preprint arXiv:2502.18965},
  year={2025}
}

@inproceedings{qarm,
  title={Qarm: Quantitative alignment multi-modal recommendation at kuaishou},
  author={Luo, Xinchen and Cao, Jiangxia and Sun, Tianyu and Yu, Jinkai and Huang, Rui and Yuan, Wei and Lin, Hezheng and Zheng, Yichen and Wang, Shiyao and Hu, Qigen and others},
  booktitle={Proceedings of the 34th ACM International Conference on Information and Knowledge Management},
  pages={5915--5922},
  year={2025}
}

@inproceedings{mt5,
  title={mT5: A massively multilingual pre-trained text-to-text transformer},
  author={Xue, Linting and Constant, Noah and Roberts, Adam and Kale, Mihir and Al-Rfou, Rami and Siddhant, Aditya and Barua, Aditya and Raffel, Colin},
  booktitle={Proceedings of the 2021 conference of the North American chapter of the association for computational linguistics: Human language technologies},
  pages={483--498},
  year={2021}
}

@article{adesde,
  title={Exploring dual encoder architectures for question answering},
  author={Dong, Zhe and Ni, Jianmo and Bikel, Daniel M and Alfonseca, Enrique and Wang, Yuan and Qu, Chen and Zitouni, Imed},
  journal={arXiv preprint arXiv:2204.07120},
  year={2022}
}

@incollection{dcn1,
  title={Deep \& cross network for ad click predictions},
  author={Wang, Ruoxi and Fu, Bin and Fu, Gang and Wang, Mingliang},
  booktitle={Proceedings of the ADKDD'17},
  pages={1--7},
  year={2017}
}

@inproceedings{dcn2,
  title={Dcn v2: Improved deep \& cross network and practical lessons for web-scale learning to rank systems},
  author={Wang, Ruoxi and Shivanna, Rakesh and Cheng, Derek and Jain, Sagar and Lin, Dong and Hong, Lichan and Chi, Ed},
  booktitle={Proceedings of the web conference 2021},
  pages={1785--1797},
  year={2021}
}

@inproceedings{dnn,
  title={Deep neural networks for youtube recommendations},
  author={Covington, Paul and Adams, Jay and Sargin, Emre},
  booktitle={Proceedings of the 10th ACM conference on recommender systems},
  pages={191--198},
  year={2016}
}

@inproceedings{din,
  title={Deep interest network for click-through rate prediction},
  author={Zhou, Guorui and Zhu, Xiaoqiang and Song, Chenru and Fan, Ying and Zhu, Han and Ma, Xiao and Yan, Yanghui and Jin, Junqi and Li, Han and Gai, Kun},
  booktitle={Proceedings of the 24th ACM SIGKDD international conference on knowledge discovery \& data mining},
  pages={1059--1068},
  year={2018}
}

@inproceedings{widedeep,
  title={Wide \& deep learning for recommender systems},
  author={Cheng, Heng-Tze and Koc, Levent and Harmsen, Jeremiah and Shaked, Tal and Chandra, Tushar and Aradhye, Hrishi and Anderson, Glen and Corrado, Greg and Chai, Wei and Ispir, Mustafa and others},
  booktitle={Proceedings of the 1st workshop on deep learning for recommender systems},
  pages={7--10},
  year={2016}
}

@article{qwen3,
  title={Qwen3 technical report},
  author={Yang, An and Li, Anfeng and Yang, Baosong and Zhang, Beichen and Hui, Binyuan and Zheng, Bo and Yu, Bowen and Gao, Chang and Huang, Chengen and Lv, Chenxu and others},
  journal={arXiv preprint arXiv:2505.09388},
  year={2025}
}

@article{llama3,
  title={The llama 3 herd of models},
  author={Dubey, Abhimanyu and Jauhri, Abhinav and Pandey, Abhinav and Kadian, Abhishek and Al-Dahle, Ahmad and Letman, Aiesha and Mathur, Akhil and Schelten, Alan and Yang, Amy and Fan, Angela and others},
  journal={arXiv e-prints},
  pages={arXiv--2407},
  year={2024}
}

@article{lora,
  title={Lora: Low-rank adaptation of large language models.},
  author={Hu, Edward J and Shen, Yelong and Wallis, Phillip and Allen-Zhu, Zeyuan and Li, Yuanzhi and Wang, Shean and Wang, Lu and Chen, Weizhu and others},
  journal={ICLR},
  volume={1},
  number={2},
  pages={3},
  year={2022}
}

@inproceedings{bert,
  title={Bert: Pre-training of deep bidirectional transformers for language understanding},
  author={Devlin, Jacob and Chang, Ming-Wei and Lee, Kenton and Toutanova, Kristina},
  booktitle={Proceedings of the 2019 conference of the North American chapter of the association for computational linguistics: human language technologies, volume 1 (long and short papers)},
  pages={4171--4186},
  year={2019}
}

@inproceedings{bge,
  title={C-pack: Packed resources for general chinese embeddings},
  author={Xiao, Shitao and Liu, Zheng and Zhang, Peitian and Muennighoff, Niklas and Lian, Defu and Nie, Jian-Yun},
  booktitle={Proceedings of the 47th international ACM SIGIR conference on research and development in information retrieval},
  pages={641--649},
  year={2024}
}

@inproceedings{xlm,
  title={Unsupervised cross-lingual representation learning at scale},
  author={Conneau, Alexis and Khandelwal, Kartikay and Goyal, Naman and Chaudhary, Vishrav and Wenzek, Guillaume and Guzm{\'a}n, Francisco and Grave, Edouard and Ott, Myle and Zettlemoyer, Luke and Stoyanov, Veselin},
  booktitle={Proceedings of the 58th annual meeting of the association for computational linguistics},
  pages={8440--8451},
  year={2020}
}

@inproceedings{lcrec,
  title={Adapting large language models by integrating collaborative semantics for recommendation},
  author={Zheng, Bowen and Hou, Yupeng and Lu, Hongyu and Chen, Yu and Zhao, Wayne Xin and Chen, Ming and Wen, Ji-Rong},
  booktitle={2024 IEEE 40th International Conference on Data Engineering (ICDE)},
  pages={1435--1448},
  year={2024},
  organization={IEEE}
}

@article{wukong,
  title={Wukong: Towards a scaling law for large-scale recommendation},
  author={Zhang, Buyun and Luo, Liang and Chen, Yuxin and Nie, Jade and Liu, Xi and Guo, Daifeng and Zhao, Yanli and Li, Shen and Hao, Yuchen and Yao, Yantao and others},
  journal={arXiv preprint arXiv:2403.02545},
  year={2024}
}

@inproceedings{gdcn,
  title={Towards deeper, lighter and interpretable cross network for ctr prediction},
  author={Wang, Fangye and Gu, Hansu and Li, Dongsheng and Lu, Tun and Zhang, Peng and Gu, Ning},
  booktitle={Proceedings of the 32nd ACM international conference on information and knowledge management},
  pages={2523--2533},
  year={2023}
}

@inproceedings{finalmlp,
  title={FinalMLP: an enhanced two-stream MLP model for CTR prediction},
  author={Mao, Kelong and Zhu, Jieming and Su, Liangcai and Cai, Guohao and Li, Yuru and Dong, Zhenhua},
  booktitle={Proceedings of the AAAI conference on artificial intelligence},
  volume={37},
  number={4},
  pages={4552--4560},
  year={2023}
}
